\documentclass[runningheads]{llncs}
\usepackage{graphicx}
\usepackage{amsmath}
\usepackage{amssymb}
\usepackage{lipsum}
\newcommand{\R}{\mathbb{R}}
\usepackage{url}
\begin{document}
\title{BabyNet: Residual Transformer Module for Birth Weight Prediction on Fetal Ultrasound Video}
\titlerunning{BabyNet: Birth Weight Prediction from Fetal US}

\author{Szymon Płotka\inst{1, 2} \and
Michal K. Grzeszczyk\inst{1} \and
Robert Brawura-Biskupski-Samaha\inst{3} \and
Paweł Gutaj\inst{4} \and
Michał Lipa\inst{5} \and
Tomasz Trzciński \inst{6} \and
Arkadiusz Sitek \inst{1}}
\authorrunning{S. Płotka et al.}

\newcommand\blfootnote[1]{%
  \begingroup
  \renewcommand\thefootnote{}\footnote{#1}%
  \addtocounter{footnote}{-1}%
  \endgroup
}

\institute{Sano Centre for Computational Medicine, Cracow, Poland \email{s.plotka@sanoscience.org} \and
Informatics Institute, University of Amsterdam, Amsterdam, The Netherlands \and
The Medical Centre of Postgraduate Education, Warsaw, Poland \and
Poznan University of Medical Sciences, Poznan, Poland \and
Medical University of Warsaw, Warsaw, Poland \and
Warsaw University of Technology, Warsaw, Poland}

\maketitle              
\blfootnote{S. Płotka and M. K. Grzeszczyk -- Authors contributed equally.}

\begin{abstract}

Predicting fetal weight at birth is an important aspect of perinatal care, particularly in the context of antenatal management, which includes the planned timing and the mode of delivery. Accurate prediction of weight using prenatal ultrasound is challenging as it requires images of specific fetal body parts during advanced pregnancy which is difficult to capture due to poor quality of images caused by the lack of amniotic fluid. As a consequence, predictions which rely on standard methods often suffer from significant errors. In this paper we propose the Residual Transformer Module which extends a 3D ResNet-based network for analysis of $2D+t$ spatio-temporal ultrasound video scans. Our end-to-end method, called BabyNet, automatically predicts fetal birth weight based on fetal ultrasound video scans. We evaluate BabyNet using a dedicated clinical set comprising 225 2D fetal ultrasound videos of pregnancies from 75 patients performed one day prior to delivery. Experimental results show that BabyNet outperforms several state-of-the-art methods and estimates the weight at birth with accuracy comparable to human experts. Furthermore, combining estimates provided by human experts with those computed by BabyNet yields the best results, outperforming either of other methods by a significant margin. The source code of BabyNet is available at \url{https://github.com/SanoScience/BabyNet}.

\keywords{Deep learning \and Fetal birth weight \and Transformer}
\end{abstract}
\section{Introduction}

Fetal birth weight (FBW) is a significant indicator of perinatal health prognosis. Accurate prediction of FBW, as well as gestational age, complications in pregnancy, and maternal physical parameters are critical in determining the best method of delivery (natural or Cesarean). These factors are widely used as a part of the hospital admission procedure in the World \cite{pressman2000prediction}. However, FBW prediction is a challenging task, requiring highly visible fetal body standard planes, which can only be identified by experienced sonographers. Unfortunately, weight predictions provided by experienced sonographers are often imprecise, with up to 10\% mean absolute percentage errors. Currently, FBW is estimated on the basis of fetal biometric measurements of body organs -- head circumference (HC), biparietal diameter (BPD), abdominal circumference (AC), femur length (FL), which are used as the input to heuristic formulae \cite{hadlock1985estimation}, \cite{milner2018accuracy}.

In recent years, machine learning-based methods have been proposed as a possible means of automating FBW prediction. Lu et al. \cite{lu2020prediction}, \cite{lu2019ensemble} presents a solution based on an ensemble model consisting of Random Forest, XGBoost and LightGBM algorithms. Tao et al. \cite{tao2021fetal} use a hybrid-LSTM network model \cite{xingjian2015convolutional} for temporal data analysis. Convolutional neural network (CNN)-based models are also proposed to estimate fetal weight based on ultrasound images \cite{bano2021autofb}, \cite{feng2019fetal} or videos \cite{plotka2021fetalnet}, \cite{plotka2022deep}. However, such methods do not rely on the true FBW as the ground truth, but instead predict it through heuristic formulae using estimated fetal body-part biometrics, which is prone to errors. 

Recently, Transformers \cite{vaswani2017attention} have been proposed as an alternative architecture to CNNs, and have achieved competitive performance for many computer vision tasks e.g. Vision Transformer (ViT) for image classification \cite{dosovitskiy2021an} or Video Vision Transformer (ViViT) for video recognition \cite{Arnab_2021_ICCV}. Transformers utilize the Multi-Head Self-Attention (MHSA) mechanism to learn the global context between input sequence elements. Unfortunately, due to their high computational complexity, Transformers require a large amount of training data and long training times. Many methods have been developed to bridge the gap between sample-efficient learning with a high inductive bias of CNNs and performance but data-inefficient Transformers. Hybrid models utilizing CNN layers and Transformer blocks have also been introduced \cite{d2021convit}, \cite{liu2021transformer}, \cite{reynaud2021ultrasound}.

In this paper we utilize Transformers for direct estimations of fetal weights from US videos. We implement this solution as an extension of a 3D ResNet-based network \cite{tran2018closer} with a Residual Transformer Module (RTM) called BabyNet. The RTM allows local and global feature representation through residual connections and utilization of convolutional layers. This representation is refined through the global self-attention mechanism included inside RTM. BabyNet is a hybrid neural network that efficiently bridges CNNs and Transformers for $2D+t$ spatio-temporal ultrasound video scans analysis to directly predict fetal birth weight.
The main contribution of our work is as follows: (1) We provide an end-to-end method for birth weight estimation based directly on fetal ultrasound video scans, (2) We introduce a novel Residual Transformer Module by adding temporal position encoding to 3D MHSA in 3D ResNet-based neural network, (3) To the best of our knowledge, BabyNet is the first framework to automate fetal birth weight prediction on fetal ultrasound video scans trained and validated with data acquired one day prior to delivery.

\begin{figure}[ht!]
    \centering
    \includegraphics[width=12cm]{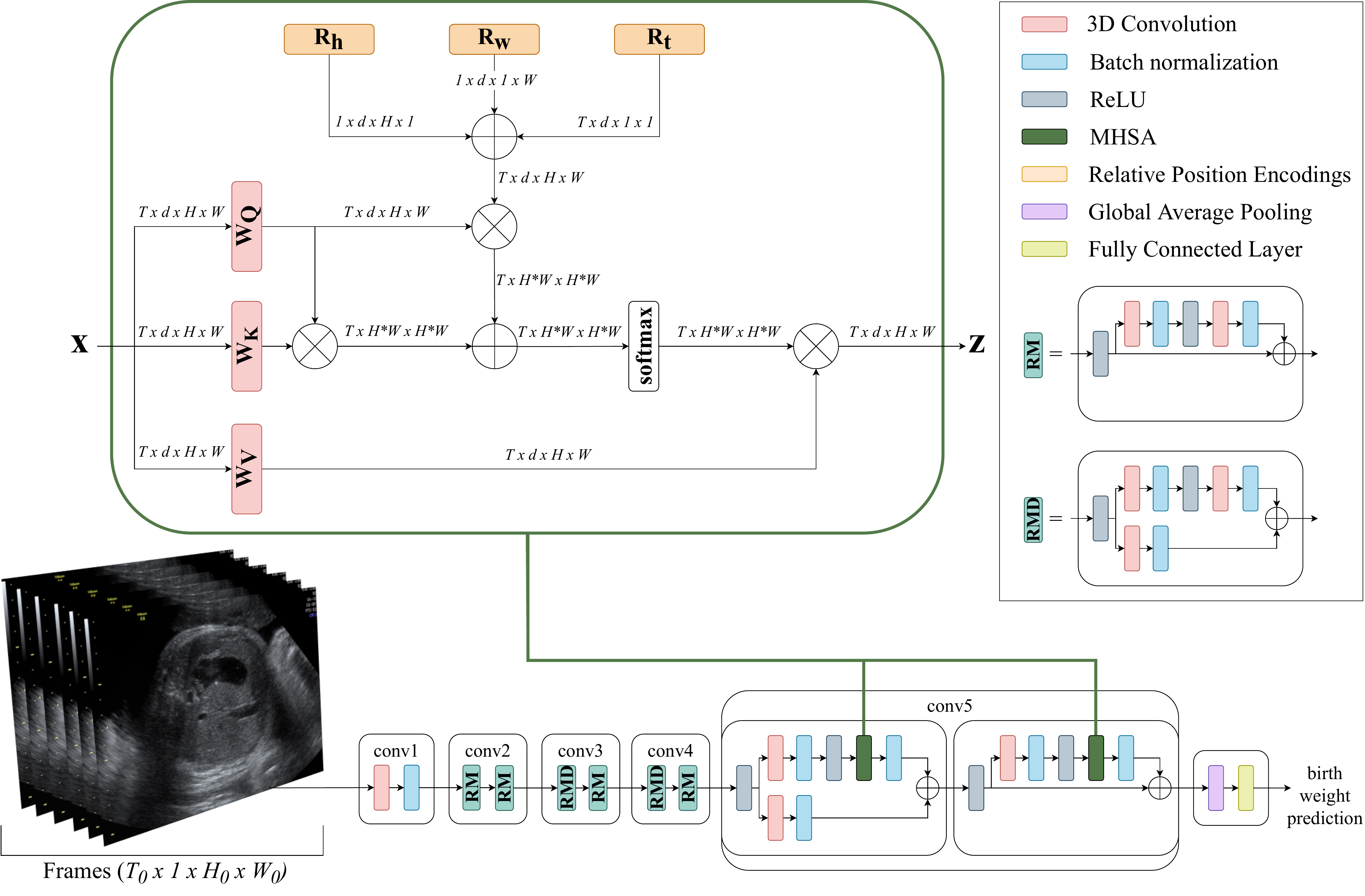}
    \caption{The overview of our proposed BabyNet method for birth weight estimation directly from fetal US video scans. In BabyNet, we replace two Residual Modules of 3D ResNet-18 with two Residual Transformer Modules (RTM) containing 3D Multi-Head Self-Attention (MHSA) with Relative Positional Encoding (RPE). RPE is calculated as the sum of height ($R_h$), width ($R_w$) and temporal ($R_t$) position encodings. For clarity, only one attention head is presented in the image. The network takes 16 consecutive frames as the input to make a single-segment prediction. All frames for a given patient are divided into non-overlapping 16-frame segments and a patient-level prediction is obtained by averaging all segment predictions.}
    \label{fig:babynet}
\end{figure}

\section{Method}
\label{sec:method}

The overview of our method for end-to-end FBW prediction is presented in Fig. \ref{fig:babynet}. We use 3D ResNet-18 for high-level US feature extraction. The RTM is designed to learn local and global feature representation with 3D Multi-Head Self-Attention mechanism and convolutional layers. We replace the last two residual modules of ResNet with RTMs. 

\subsection{Feature Extraction} 

We employ 3D ResNet-18 \cite{tran2018closer} as the base network to extract high-level $2D+t$ spatio-temporal US feature representations. The initial input to the network is US video sequence $S_{US} \in \R^{T_{0} \times 1 \times H_{0} \times W_{0}}$ of height $H_0$, width $W_0$ and frame number $T_0$. It is transformed via convolutional residual modules to a low-resolution feature map sequence $S^{'}_{US} \in \R^{T_{1} \times D_{1} \times H_{1} \times W_{1}}$, where $T_{1} = T_{0}/4$, $D_{1} = 512$, $H_{1} = H_{0}/8$, and $W_{1} = W_{0}/8$. Multi-channel, low-resolution feature map sequences are fed to the RTM. 

\subsection{Residual Transformer Module} Residual modules are constructed from a layer followed by a rectified linear unit (ReLU) and Batch Normalization. This structure is repeated two or three times with a skip connection of the input added to the output of the previous layers \cite{he2016deep}. To include global low-resolution feature map context processing via a self-attention mechanism we design RTM in a similar manner to BoT \cite{srinivas2021bottleneck}. Our RTM extends BoT to 3D space by adding temporal position encoding \cite{shaw2018self} to 3D Multi-Head Self-Attention. BoT utilizes MHSA instead of $3 \times 3$ convolution in the residual bottleneck module, created to decrease the computational complexity in deeper ResNet architectures. 3D ResNets are often shallower and do not contain Bottleneck blocks. Thus, to utilize the self-attention mechanism in shallower ResNets we replace the last convolutional layer in the residual module with MHSA, and define RTM as: 

\begin{equation}
y = BN\left(MHSA(\sigma(BN(Conv(x)\right) + x
\end{equation}
where $x$ and $y$ are input and output of the RTM respectively, $Conv$ denotes the convolutional layer, $BN$ is Batch Normalization and $\sigma$ stands for ReLU.

\subsection{3D Multi-Head Self-Attention} To learn multiple attention representations at different positions, instead of performing a single attention, many self-attention heads (Multi-Head Self-Attention) are jointly trained with their outputs concatenated \cite{vaswani2017attention}. Since such operation is permutation-invariant, positional encoding \textit{r} needs to be added to include positional information. Depending on the application, absolute (e.g. sinusoidal) or relative positional encodings (RPE) \cite{shaw2018self}, recently identified as a better fit for vision tasks \cite{wu2021rethinking}, can be used. To process $2D + t$ US videos with MHSA we add temporal positional encoding to the 2D RPE and compute positional encoding \textit{r} as the sum of $R_h\in\mathbb{R}^{1{\times}D{\times}H{\times}1}$, $R_w\in\mathbb{R}^{1{\times}D{\times}1{\times}W}$ and $R_t\in\mathbb{R}^{T{\times}D{\times}1{\times}1}$, the height, width and temporal positional encodings respectively. Finally, we compute the 3D MHSA output of $S^{''}_{US} \in \R^{T \times D \times H \times W}$ input as:
\begin{equation}
MHSA\left(S^{''}_{US}\right) = concat\left[softmax\left(\frac{Q_i(K_i+r)^T}{\sqrt{d}}\right)V_i\right]
\end{equation}
where $T=\frac{T_{1}}{2}$, $D=D_{1}$, $H=\frac{H_{1}}{2}$, $W=\frac{W_{1}}{2}$, $Q_i$, $K_i$, $V_i$ are queries, keys and values for the \textit{i}th attention head calculated from $W_Q(S^{''}_{US})$, $W_K(S^{''}_{US})$ and $W_V(S^{''}_{US})$ $1 \times 1 \times 1$ 3D convolutions performed over input \textit{$S^{''}_{US}$} and $d$ is $D$ divided by the number of heads.
\section{Experiments}
\label{sec:experiments}
In this section, we describe our dataset and present architectural details of BabyNet. We compare BabyNet's performance with other $2D+t$ spatio-temporal video analysis methods and with results obtained from clinicians. We show, through an ablation study, the importance of BabyNet components that have been added or replaced in 3D ResNet-18.

\begin{figure*}[t!]
\minipage{0.33\textwidth}
  \includegraphics[width=\linewidth, height=3cm]{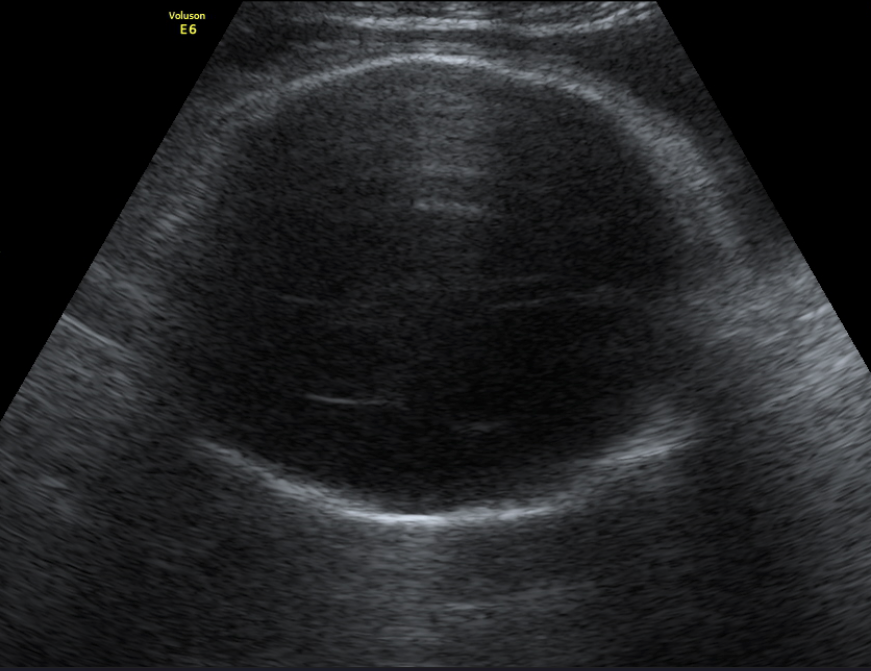}
\endminipage\hfill
\minipage{0.33\textwidth}
  \includegraphics[width=\linewidth, height=3cm]{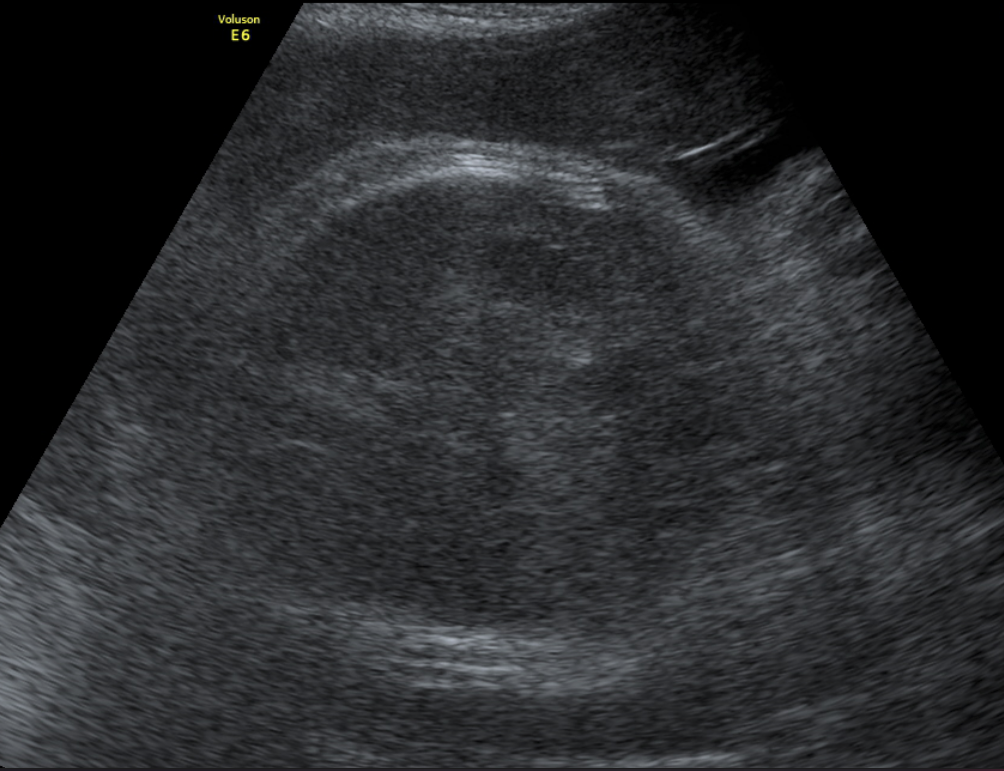}
\endminipage\hfill
\minipage{0.33\textwidth}%
  \includegraphics[width=\linewidth, height=3cm]{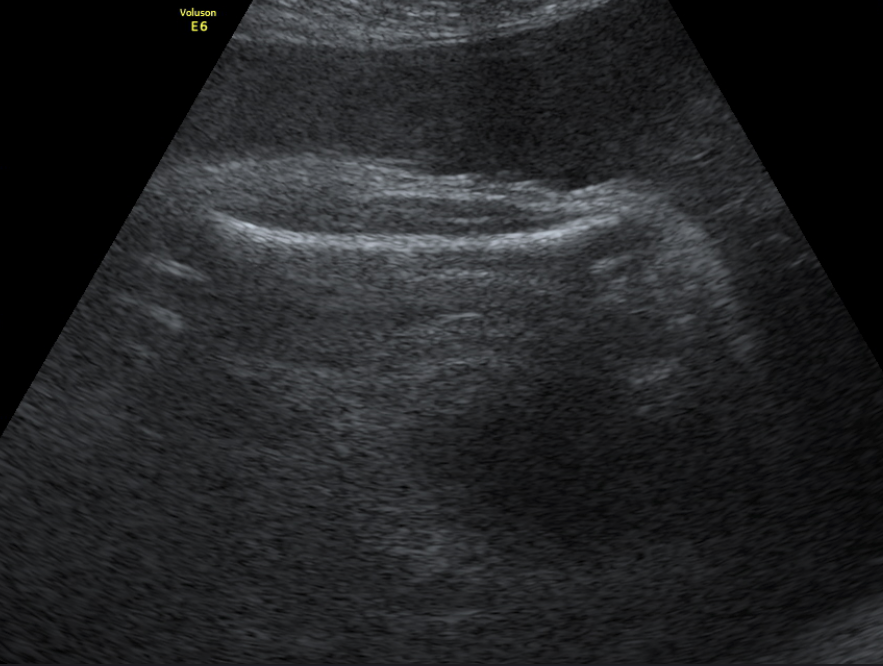}
\endminipage
\caption{Sample US frames extracted from fetal US videos. The frames show the fetal body part standard planes of the head, abdomen and femur respectively, going from left to right. Images obtained several hours before delivery are of lower quality than at earlier stages of pregnancy due to the lack of amniotic fluid.}
\label{fig:stplanes}
\end{figure*}

\subsubsection{Dataset and Pre-processing.} Ethical Committee approval was obtained for all subjects enrolled in the study. The dataset consists of 225 2D fetal ultrasound video scans in standard plane view of fetal head, abdomen, and femur. The multi-centre dataset was obtained from 75 pregnant women aged 21 to 42 and acquired through routine US examinations less than 24 hours prior to delivery. The data was acquired by three experienced sonographers using GE Voluson E6 and S10 devices. Each US video scan is stored in DICOM file format, captured in two resolutions: $960 \times 720$ and $852 \times 1136$ pixels. The number of frames is between 463 and 1448, with a mean of $852$. The US videos were obtained in sector scan sweep mode with frame per second (FPS) between 24 and 37. For each video, we resample pixel spacing to $0.2 \times 0.2$ mm. As the ground truth, we use the true fetal weight measured at birth. The ground truth values were between $2085$ and $4995$, with a mean of $3454$ grams [g].

\subsubsection{Implementation Details.} We adopt 3D ResNet-18 \cite{tran2018closer} as our base neural network. Table \ref{tab:babynetarchitecture} presents the architectural details of BabyNet, as compared to 3D ResNet-18. BabyNet comprises a 3D convolutional stem followed by \textit{conv} stages: three with two residual modules each, and one final stage implemented with two RTMs. The output of the final RTM is global average pooled (GAP) and fed to the fully-connected (FC) layer with one neuron (512 input weights) for fetal birth weight prediction. We implement our model with PyTorch and train it using an NVIDIA RTX 2080 Ti 24GB GPU with a mini-batch size of 2 and an initial learning rate of $1 \times 10^{-4}$ with a step decay by a factor of $g = 0.1$ every $160^{th}$ epochs until convergence over 200 epochs. To minimize the Mean Squared Error (MSE) loss function, we employ an ADAM optimizer with $1 \times 10^{-4}$ weight decay. During training, we apply data augmentation including rotate ($\pm$ 25$^{\circ}$), random brightness and contrast, horizontal flip, image compression and blur for each mini-batch. We retain height and width ratio and resize video frames to $64 \times 64$ ($H_{0} \times W_{0}$) with padding. The number of attention heads is empirically set to 4, while the temporal sequence length $T_{0}$ is 16. Thus, BabyNet transforms US input sequence $S_{US} \in \R^{16 \times 1 \times 64 \times 64}$ to the output $O_{S_{US}} \in \R^{1}$ of predicted fetal birth weights. We perform 5-fold cross-validation (CV) to compare and verify the robustness of the regression algorithm. We ensure that data from a single patient appears only in a single fold. 

\bgroup
\def\arraystretch{1.5}%
\begin{table}[t!]
    \caption{Comparison of ResNet3D-18 and BabyNet architectures. We replace the last two residual modules of 3D ResNet-18 with two Residual Transformer Modules containing a 3D MHSA instead of the second $3 \times 3$ 3D convolution.}\label{tab:babynetarchitecture}
    \begin{center}
        \begin{tabular}{c|c|c|c}
            Stage name & Output size & 3D ResNet-18 & BabyNet\\
            \hline
            conv1 &  \textit{$T_{0}\times\frac{H_{0}}{2}\times\frac{W_{0}}{2}$} & \multicolumn{2}{|c}{$3\times7\times7, 64,$ stride $1\times2\times2$}\\ \hline
            conv2 &   \textit{$T_{0}\times\frac{H_{0}}{2}\times\frac{W_{0}}{2}$} & 
                \Bigg[\begin{tabular}{@{}c@{}} $3\times3\times3, 64$ \\ $3\times3\times3, 64$  \end{tabular} \Bigg] $\times2$ &
                \Bigg[\begin{tabular}{@{}c@{}} $3\times3\times3, 64$ \\ $3\times3\times3, 64$  \end{tabular} \Bigg] $\times2$ \\ \hline
            conv3 &  \textit{$\frac{T_{0}}{2}\times\frac{H_{0}}{4}\times\frac{W_{0}}{4}$} & 
                \Bigg[\begin{tabular}{@{}c@{}} $3\times3\times3, 128$ \\ $3\times3\times3, 128$  \end{tabular}\Bigg] $\times2$ & 
                \Bigg[\begin{tabular}{@{}c@{}} $3\times3\times3, 128$ \\ $3\times3\times3, 128$  \end{tabular} \Bigg] $\times2$ \\ \hline
            conv4 &  \textit{$\frac{T_{0}}{4}\times\frac{H_{0}}{8}\times\frac{W_{0}}{8}$} & 
                \Bigg[\begin{tabular}{@{}c@{}} $3\times3\times3, 256$ \\ $3\times3\times3, 256$  \end{tabular}\Bigg] $\times2$ & 
                \Bigg[\begin{tabular}{@{}c@{}} $3\times3\times3, 256$ \\ $3\times3\times3, 256$  \end{tabular}\Bigg] $\times2$ \\ \hline
            conv5 &  \textit{$\frac{T_{0}}{8}\times\frac{H_{0}}{16}\times\frac{W_{0}}{16}$} & 
                \Bigg[\begin{tabular}{@{}c@{}} $3\times3\times3, 512$ \\ $3\times3\times3, 512$  \end{tabular}\Bigg] $\times2$ & 
                \Bigg[$\overbrace{\begin{tabular}{@{}c@{}} $3\times3\times3, 512$ \\ \textbf{MHSA}  \end{tabular}}^{RTM}$\Bigg] $\times2$ \\ \hline
            & \textit{$1\times1\times1$}& \multicolumn{2}{|c}{Global Avg Pooling, FC layer} \\
        \end{tabular}
    \end{center}
\end{table}
\egroup

\subsubsection{Evaluation Metrics.} As measurement metrics, we use Root Mean Square Error (RMSE), Mean Absolute Error (MAE), and Mean Absolute Percentage Error (MAPE) to evaluate the regression performance.

\subsubsection{Comparison with Clinicians and State-of-the-art Algorithms.} We compare BabyNet with several $2D+t$ spatio-temporal video analysis methods. In particular, we compare it with results obtained by clinicians in \cite{sherman1998comparison} as well as results obtained by clinicians for the dataset used in this work. We also present results for Video Vision Transformer (ViViT) \cite{Arnab_2021_ICCV} and test the hybrid approach of 2D ResNet-50 as a convolutional feature extractor (without GAP and FC layers) to ViViT network. Finally, we utilize a vanilla 3D ResNet-18 \cite{tran2018closer}. We train all models in the same fashion as BabyNet.

Table \ref{tab:resultssota} presents a comparison of 5-fold CV results for all tested methods. Results for machine learning methods are out-of-fold predictions. Combination of estimations performed by clinicians with estimations provided by BabyNet is the most accurate, with MAE of $180 \pm 156$ (max p-value $< 0.001$), RMSE of $237 \pm 145$ (max p-value $< 0.001$), and MAPE of $5.2 \pm 4.6$ (max p-value $< 0.001$). Max p-value is the maximum paired dual sided p-value computed for results of the "Clinicians (this work) \& BabyNet" method and other methods listed in Table~\ref{tab:resultssota}.

We did not detect a statistically significant difference between the performance of clinicians measured in \cite{sherman1998comparison} and our algorithm (p-value = 0.6). Estimations provided by clinicians in our study seem to be better than those provided by clinicians in \cite{sherman1998comparison} (p-value = 0.04) and BabyNet (p-value = 0.07). Out of all neural networks investigated in this work, the hybrid approach of utilizing 3D convolutions and 3D MHSA within RTM as a part of 3D ResNet-18 outperforms other methods based on plain CNNs, plain Transformer or CNN+Transformer networks. 

We noted that the best results were obtained by averaging estimations provided by clinicians and by BabyNet. The performance of the ensemble of clinicians \& BabyNet was better by 18\% compared to clinicians alone in terms of mMAPE, which is a clear indication of added value and potential clinical benefits of BabyNet.  

\bgroup
\def\arraystretch{1.5}%
\begin{table}[t!]
    \caption{Five-fold cross-validation results and comparison of state-of-the-art methods. The mean of Mean Absolute Error (MAE), Root Mean Square Error (RMSE) and Mean Absolute Percentage Error (MAPE) across all folds are reported.}\label{tab:resultssota}
    \begin{center}
        \begin{tabular}{c|c|c|c}
            Method & mMAE [g] & mRMSE [g] & mMAPE [\%]\\
            \hline
            Clinicians (from \cite{sherman1998comparison}) & - & - & $7.9 \pm 6.8$\\ \hline
            Clinicians (this work)  & $213 \pm 155$ & $264 \pm 158$ & $6.3 \pm 4.8$\\ \hline
            ViViT \cite{Arnab_2021_ICCV} & $361 \pm 244$ & $444 \pm 230$ & $10.6 \pm 7.3$ \\ \hline
            2D ResNet + ViViT & $344 \pm 241$ & $426 \pm 226$ & $10.3 \pm 7.2$ \\ \hline
            3D ResNet-18 \cite{tran2018closer} & $328 \pm 234$ & $421 \pm 225$ & $10.1 \pm 7.1$\\ \hline
            BabyNet & $254 \pm 230$  & $341 \pm 215$ & $7.5 \pm 6.6$ \\ \hline
             \textbf{ Clinicians (this work) \& BabyNet}  & $\mathbf{180 \pm 156}$ & $\mathbf{237 \pm 145}$ & $\mathbf{5.2 \pm 4.6}$\\             
        \end{tabular}
    \end{center}
\end{table}
\egroup

\subsubsection{Ablation study.} We conducted an ablation study to show the effectiveness of novel components within BabyNet. In this experiment, we employ 3D ResNet-18 as the base neural network for $2D+t$ spatio-temporal US video scan analysis. To learn multiple relationships and enable capture of richer interpretations of the US video sequence, we integrate CNN and Transformer by swapping the last convolutional layer in the residual module for MHSA. 
To further enhance $2D+t$ spatio-temporal feature representation in space and time, we add temporal position encoding (TPE).
Table \ref{tab:ablationstudy} demonstrates that the combination of CNN with a Transformer-based module, MHSA and temporal position encoding  improves  performance of the weight-estimation task directly from US video scan.

\bgroup
\def\arraystretch{1.5}%
\begin{table}[t!]
    \caption{Ablation study.}
    \begin{center}
        \begin{tabular}{c|c|c|c}
            Method & mMAE [g] $\downarrow$ & mRMSE [g] $\downarrow$  & mMAPE [\%] $\downarrow$\\
            \hline
            3D ResNet-18 (base) & $328 \pm 234$ & $421 \pm 225$ & $10.1 \pm 7.1$\\ \hline
            + RTM & $277 \pm 228$ & $374 \pm 221$ & $8.1 \pm 7.0$\\ \hline
            \textbf{+ RTM + TPE (ours)} & $\mathbf{254 \pm 230}$  & $\mathbf{341 \pm 215}$ & $\mathbf{7.5 \pm 6.6}$ \\
        \end{tabular}
    \end{center}
    \label{tab:ablationstudy}
\end{table}
\egroup

\section{Discussion}
\label{sec:discussion}

In this work we were not able to match the performance of the clinicians in estimating fetal weight (mMAPE 7.5\% vs. 6.3\%); however, clinicians who worked with us and provided measurements are top experts in performing biometric measurements. On the other hand, we were able to match the performance of clinicians reported in \cite{sherman1998comparison} (7.5\% vs 7.9\% p-value=0.6). The training data set was relatively small and we expect to significantly improve the performance of BabyNet by using more data in future work. 

The method presented here can be characterized as end-to-end. Due to $2D+t$ spatio-temporal feature processing it does not require standard plane detection which substantially reduces the workload involved in performing the estimation of FBW. In clinical practice, BabyNet can be used as an aid for clinicians in their decision-making process regarding the type of delivery. According to literature \cite{scioscia2008estimation}, \cite{sherman1998comparison} the heavier the child, the greater the likelihood of Cesarean delivery. Serious complications may arise when a heavy child's FBW is misjudged. Under these circumstances, if natural delivery is decided upon severe complications for both mother and child may arise.

This work has certain limitations. A relatively small number of patients was used in the study, which can affect the accuracy and generalization of results. A related issue is that the patient population is limited and we do not know if BabyNet would work on a different population (e.g. different race). The algorithm is trained and evaluated on short clips of US videos recorded by clinicians. To operate in a clinical setting, further effort would be needed to create a system that extracts appropriate clips for BabyNet analysis.

\section{Conclusions}
\label{sec:conclusions}

In this paper we presented an extension of the 3D ResNet-based network with a Residual Transformer Module (RTM), named BabyNet, for $2D+t$ spatio-temporal fetal ultrasound video scan analysis. The proposed framework is an end-to-end method that automatically performs fetal birth weight prediction. This is done without the need for finding standard planes in ultrasound video scans, which are required in the classical method of estimating fetal weight. Combining classical and BabyNet estimations provides the best results, significantly outperforming top expert clinicians who use available commercial tools. Our method has the potential to help clinicians select -- on the basis of US examination -- the type of delivery which is safest for the mother and the child. Future work includes testing BabyNet on external datasets which are preferably acquired using different devices and by operators with different levels of experience. Moreover, we plan to use multimodal data -- combine the fetal US video and clinical data to improve the performance and robustness of the model.

\section*{Acknowledgements}

This work is supported by the European Union’s Horizon 2020 research and innovation programme under grant agreement Sano No 857533 and the International Research Agendas programme of the Foundation for Polish Science, co-financed by the European Union under the European Regional Development Fund. We would like to thank Piotr Nowakowski for his assistance with proofreading the manuscript.

%
%
%
%
\bibliographystyle{splncs04}
\bibliography{references}
\end{document}